\documentclass{aastex}
\usepackage{emulateapj5,times,mathptm}
\singlespace
\slugcomment{Accepted for the publication in the {\it Astrophysical Journal}}
\def\farcm{\hbox{$.\mkern-4mu^\prime$}}

\def\la{\mathrel{\hbox{\rlap{\hbox{\lower4pt\hbox{$\sim$}}}\hbox{$<$}}}}
\def\ga{\mathrel{\hbox{\rlap{\hbox{\lower4pt\hbox{$\sim$}}}\hbox{$>$}}}}

\shortauthors{Park}
\shorttitle{Galactic center}

\begin{document}
\title{Characteristics of Diffuse X-Ray Line Emission within 20 pc of 
the Galactic Center}
\author{Sangwook Park\altaffilmark{1,5}, Michael P. Muno\altaffilmark{2},
Frederick K. Baganoff\altaffilmark{2}, Yoshitomo Maeda\altaffilmark{3}, 
Mark Morris\altaffilmark{4}, Christian Howard\altaffilmark{4}, Mark W. 
Bautz\altaffilmark{2}, and Gordon P. Garmire\altaffilmark{1}}

\altaffiltext{1}{Department of Astronomy and Astrophysics, Pennsylvania 
State University, 525 Davey Laboratory, University Park, PA. 16802}
\altaffiltext{2}{Center for Space Research, Massachusetts Institute of
Technology, Cambridge, MA. 02139 }
\altaffiltext{3}{Institute of Space and Astronautical Science, 3-1-1 
Yoshinodai, Sagamihara, Kanagawa, 229-8510, Japan}
\altaffiltext{4}{Department of Physics and Astronomy, University of
California Los Angeles, Los Angeles, CA. 90095}
\altaffiltext{5}{park@astro.psu.edu}

\begin{abstract}

Over the last three years, the Galactic center region has been monitored 
with the {\it Chandra X-Ray Observatory}. Besides the X-ray emission from 
the target object, Sgr A*, diffuse X-ray emission was detected throughout 
most of the 17$'$ $\times$ 17$'$ field of view.                           
With 11 {\it Chandra} observations through 2002 June, the total effective 
exposure reaches $\sim$590 ks, providing significant photon statistics on 
much of the detailed structure within the faint, often filamentary,       
diffuse X-ray emission. The {\it true-color} X-ray image and the {\it 
equivalent width} images for the detected elemental species of the Galactic 
center region demonstrate that the diffuse X-ray features have a broad range 
of spatio-spectral properties. Enhancements of the low-ionization-state Fe 
line emission ($E$ $\sim$ 6.4 keV) to the northeast of Sgr A* can be 
interpreted as fluorescence within the dense interstellar medium resulting 
from irradiation by hard, external X-ray sources. They may also be explained
by emission induced by the bombardments by high energy particles on the
ambient medium, such as may accompany unresolved supernova ejecta intruding 
into dense interstellar medium. The detection of molecular cloud counterparts 
to the 6.4 keV Fe line emitting features indicates that these Fe line 
features are associated with dense Galactic center clouds and/or active 
star-forming regions, which strongly supports the proposed origins of
the X-ray reflection and/or supernova ejecta for the Fe line emission. 
We detect highly ionized S and Si lines which are generally coincident with 
the low-ionization-state Fe line emission and the dense molecular clouds 
in the northeast of Sgr A*. These hot plasmas are then
most likely produced by massive star-forming activities and/or supernova
remnants in the Galactic center. In contrast, we find that highly ionized 
He-like Fe line emission ($E$ $\sim$ 6.7 keV) is primarily distributed 
along the plane instead of being concentrated in the northeast of Sgr A*.
The implied high temperature and the relatively uniform, but, strong 
alignment along the plane are consistent with the magnetic confinement 
model, suggesting this hot gas component has been re-heated by the strong 
interstellar magnetic fields in the Galactic center to produce the observed 
He-like Fe line emission.
\end{abstract}

\keywords {Galaxy: center -- ISM: clouds -- X-rays: individual 
(Sagittarius A East, Sagittarius A*) -- X-rays: ISM}

\section {\label {sec:intro} INTRODUCTION}

The central $\sim$100 pc around the Galactic center is an extremely
complex region containing a variety of astrophysical activities: i.e., 
cold and warm molecular clouds, star clusters/formation, supernova remnants 
(SNRs), and H{\small II} regions, to name a few. Earlier Galactic plane 
surveys with the {\it Tenma} and {\it Ginga} satellites have discovered 
the highly ionized Fe emission line at $E$ $\sim$ 6.7 keV, which is strongly
enhanced toward the Galactic center region \citep{koyama86a,koyama89}.
{\it ASCA} and {\it BeppoSAX} observations have confirmed the presence
of the ionized Fe emission line as well as detecting emission lines from 
other highly ionized elemental species of Si, S, and Ar 
\citep{koyama96,sidoli99}. These results suggested that the difuse X-ray
emission in the plane and the Galactic center regions arises from thermal 
plasmas with multiple temperatures and ionization states, although the 
temperatures of the individual plasma components are inconclusive (e.g., 
Koyama et al.  1996; Kaneda et al. 1997; Valinia \& Marshall 1998; Park \& 
Ebisawa 2001). The origins of these hot plasmas have been suggested to be 
multiple supernova explosions and magnetic confinement of the interstellar 
medium (ISM) (e.g., Kaneda et al. 1997 and references therein).

Besides the apparent hot thermal components, the presence of giant 
molecular cloud complexes (e.g., Sanders et al. 1984), as well as a number 
of hard X-ray emitting point sources near the Galactic center (e.g., 
Skinner et al. 1987) has led to the prediction that fluorescent     
X-ray emission should arise where the cold, dense ISM
is illuminated by external X-ray sources \citep{sunyaev93}. This     
proposed new class of X-ray sources has been discovered with {\it ASCA} 
by detecting low-ionization-state Fe line emission (Fe{\small I} $-$       
Fe{\small XVII}) in the Galactic center region emerging at a               
characteristic energy of $E$~$\sim$~6.4 keV \citep{koyama96}.                  
A detailed analysis of this low ionization, or ``neutral'' 6.4 keV Fe 
line emission toward the Galactic giant molecular cloud Sgr B2 has 
successfully described this fluorescent X-ray emission from the 
molecular cloud in terms of an X-ray reflection nebula (XRN) model 
\citep{mura00}. Strong past activity of the supermassive 
Galactic black hole candidate Sgr A* was proposed to be the primary 
external illuminating source to produce the observed neutral Fe line 
emission from Sgr B2. A similar XRN feature has also been detected in 
another Galactic center molecular cloud, Sgr C \citep{mura01a}.            
Alternatively, such neutral Fe line emission may also be explained by 
the bombardment of energetic electrons on metal-rich SN ejecta fragments 
\citep{bykov02}. Massive stars appear to be forming at several sites in 
the central few hundred parsecs \citep{morris93,yusef02}, so the proposed 
unresolved SN ejecta scenario for the origin of the observed neutral Fe 
line emission is plausible as well. 

The X-ray line emission from the hot/cold ISM is therefore a useful tool 
to probe the origins of the complex structure of the ISM in the Galactic
center. The poor angular resolutions of previous detectors were, however, 
insufficient for such studies in complex regions of the Galactic center. 
With the superb sub-arcsecond angular resolution and good sensitivity in 
the 0.3 $-$ 8 keV band, the {\it Chandra X-Ray Observatory} provides an 
excellent opportunity to investigate the nature of the diffuse X-ray 
emission in the Galactic center regions. Preliminary results from earlier 
{\it Chandra} observations showed that the diffuse X-ray emission features 
in the Galactic center appear to have a variety of spectral and 
morphological characteristics \citep{bamba02}. The interpretations of 
these diffuse X-ray features have however been highly speculative primarily 
due to the limited photon statistics with previous data. Recently, we have 
performed deep observations of the Galactic center with {\it Chandra}. 
The exposure has reached $\sim$600 ks, which is an order of magnitude deeper 
than previous observations, so that these deep {\it Chandra} observations 
represent an unprecedented wealth of X-ray data on the Galactic center. 
In fact, these {\it Chandra} data have resolved over $\sim$2300 point 
sources within $\sim$20 pc of the Galactic center \citep{muno03a}. 
Besides these point sources, including the target object Sgr A*, the 
{\it Chandra} data show a complex structure of the diffuse X-ray emission 
within the 17$'$ $\times$ 17$'$ field of view (FOV). We here present the 
latest deep {\it Chandra} observations of the Galactic center. The high 
angular resolution {\it Chandra} data with decent photon statistics from 
the deep exposure allow us extensive imaging and spectral analysis of 
detected bright filamentary features. The observation is described in 
\S\ref{sec:obs}, and the X-ray image analysis is presented in 
\S\ref{sec:image} and \S\ref{sec:ewi}. We present the results from spectral 
analysis for some selected emission features in \S\ref{sec:spec}, and 
compare these results with molecular line emission features in 
\S\ref{sec:mc}. Some implications are then discussed in \S\ref{sec:disc}. 
A summary is presented in \S\ref{sec:summary}.

\section{\label{sec:obs} OBSERVATIONS \& DATA REDUCTION}

For the last three years, the Galactic supermassive black hole candidate
Sgr A* has been monitored, as part of Guaranteed Time Observation (GTO)    
and Guest Observer (GO) programs, with the Advanced CCD Imaging 
Spectrometer (ACIS) \citep{garmire03} on board the {\it Chandra X-Ray 
Observatory} (Table \ref{tbl:tab1}). As of 2002 June, combining 11 {\it
Chandra}/ACIS observations (except for ObsID 1561a, which was severely    
contaminated by a bright transient source within the FOV), the total 
exposure has reached $\sim$590 ks, which is the deepest ever observation 
of the Galactic center region in X-rays. We first applied the algorithm
developed by Townsley et al. (2002a) to correct the spatial and 
spectral degradation of the ACIS data caused by radiation damage, known 
as charge transfer inefficiency (CTI; Townsley et al. [2000]). The 
expected effects of the CTI correction include an increase in the number 
of detected events and improved event energies and energy resolution 
\citep{townsley00,townsley02a}. We then screened the data by status, 
grade, and the flight timeline filter. We have also removed observation 
time intervals of strong flaring in the background. All individual event 
files were then reprojected to the tangent plane at the radio position 
of Sgr A* (RA[J2000] = 17$^h$ 45$^m$ 40$^s$.0409, Dec[J2000] = 
$-$29$^{\circ}$ 00$'$ 28$\farcs$118) in order to generate the composite 
data. The detailed descriptions of these data reduction processes and 
the resulting broadband raw image from the composite data may be found 
in Muno et al. (2003a).

\section{\label{sec:image} X-Ray Images}

\begin{figure}[]
\figurenum{1}
\centerline{\includegraphics[angle=0,width=15cm]{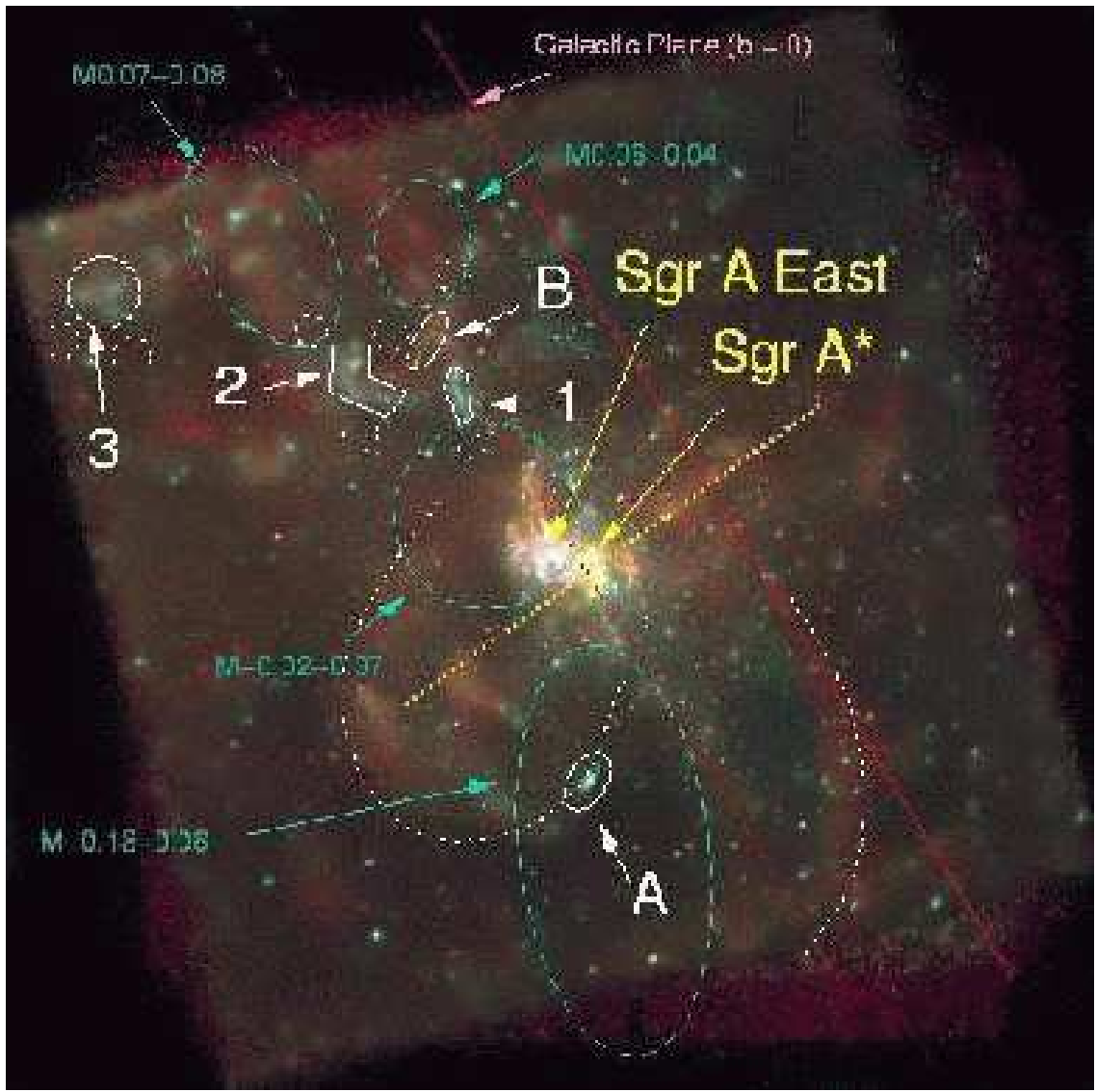}}
\figcaption[]{The exposure-corrected true color image of the Galactic 
center from the composite data of 11 {\it Chandra}/ACIS observations: red 
represents the 1.5 $-$ 4.5 keV band photons, green is 4.5 $-$ 6.0 keV, and 
blue is 6.0 $-$ 8.0 keV band. The red diagonal dashed line is the Galactic 
plane through Sgr A*. Some Galactic molecular clouds are schematically 
presented with blue dashed ellipses. Each subband image has been adaptively 
smoothed and the detected point sources have not been removed. Regions
1, 2, and 3 indicate the bright 6.4 keV Fe line emitting features.
Small dotted white circles around regions adjacent to knots 1, 2, and 3 
show where the background spectra for the Fe knots were extracted.
\label{fig:fig1}}
\end{figure}

During the monitoring observations of Sgr A*                                
(RA[J2000] = 17$^h$ 45$^m$ 40$^s$.00, Dec[J2000] = $-$29$^{\circ}$ 00$'$ 
28$\farcs$0), the 17$'$ $\times$ 17$'$ FOV of the ACIS-I array also         
imaged the interstellar surroundings of Sgr A*, which reveals the complex 
environment within $\sim$20 pc of the Galactic center with unprecedented 
detail in X-rays (Hereafter, we assume a distance of $d$ = 8 kpc to the     
Galactic center [Reid 1993]). A ``true-color'' X-ray image of the Galactic 
center region is presented in Figure \ref{fig:fig1}. Each subband image 
has been exposure-corrected utilizing the exposure map produced by Muno et 
al. (2003a), and adaptively smoothed to achieve signal-to-noise (S/N) 
ratio of 4 by using the {\it CIAO} tool {\it csmooth}. While over 
$\sim$2000 point-like sources prevail over the FOV, the point source 
contribution to the observed X-ray flux is $\sim$10\% and the bulk of the 
X-ray emission is {\it truly} diffuse \citep{muno03a}. Figure \ref{fig:fig1} 
indeed demonstrates the presence of diffuse extended X-ray emission features 
on a variety of angular scales.                                             

The target object Sgr A* is located near the center of the image where the 
bright X-ray emission from a massive star cluster within the inner parsec
of the Galactic center dominates. The bright diffuse X-ray emission to the 
immediate east of Sgr A* is the SNR Sgr A East \citep{maeda02}.   
Symmetric lobes of soft X-ray emission extended from Sgr A*, perpendicular
to the Galactic plane, are indicated with the yellow dotted line in Figure
\ref{fig:fig1}. These features may represent a bipolar outflow from
Sgr A*, which are extensively discussed by Baganoff et al. (2003) and Morris 
et al. (2003, in preparation). 

Outside of these complex central regions are larger angular scale ($\sim$a  
few arcmin) X-ray features, consisting of a few X-ray ``loops'' around 
Sgr A* (e.g., the white dotted curves in Figure \ref{fig:fig1}). The soft
nature of these loop-like features is typical of thermal shocks 
interacting with the Galactic ISM. Considering significant star-forming
activities in the Galactic center regions, these red, soft X-ray features
are perhaps caused by supernova blast waves or stellar winds/bubbles.  

We also detect a few hard X-ray knots and filaments. The bright knot to 
the south of Sgr A* (region A in Figure \ref{fig:fig1}) is the X-ray 
counterpart of the bright, nonthermal radio knot in the boundary shell 
of the SNR G359.92$-$0.09 \citep{ho85}. We have identified at least one more 
continuum dominated arc-like feature in the northeast of Sgr A* (region B 
in Figure \ref{fig:fig1}), which shows a hard photon index. The detailed 
analysis of these continuum X-ray sources are presented elsewhere 
(Baganoff et al. 2003; Sakano et al. 2003; Morris et al. 2003, in 
preparation; Yusef-Zadeh et al. 2003, in preparation). In contrast, 
some other blue, hard filaments to the northeast of Sgr A* (regions 1, 2, 
and 3 in Figure \ref{fig:fig1}) exhibit strong neutral Fe K$_{\alpha}$ line 
emission \citep{koyama96,bamba02}. The origin of these neutral Fe line 
sources has been elusive \citep{bamba02}, and we will discuss the detailed 
spectral characteristics of these neutral Fe line features in 
\S\ref{sec:spec}. 

\section{\label{sec:ewi} Equivalent Width Images}

\begin{figure}[]
\figurenum{2}
\centerline{{\includegraphics[width=0.4\textwidth]{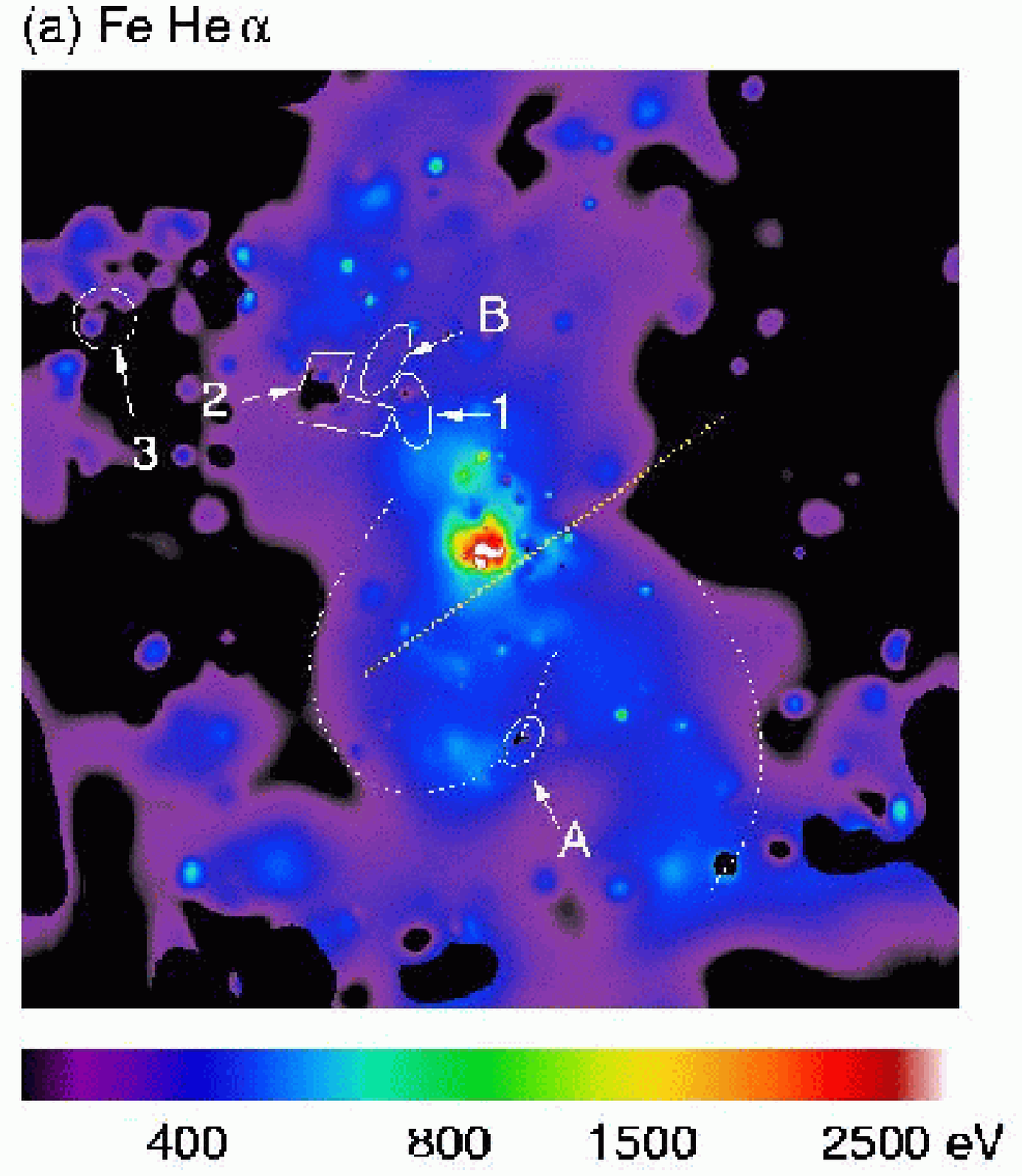}}
{\includegraphics[width=0.4\textwidth]{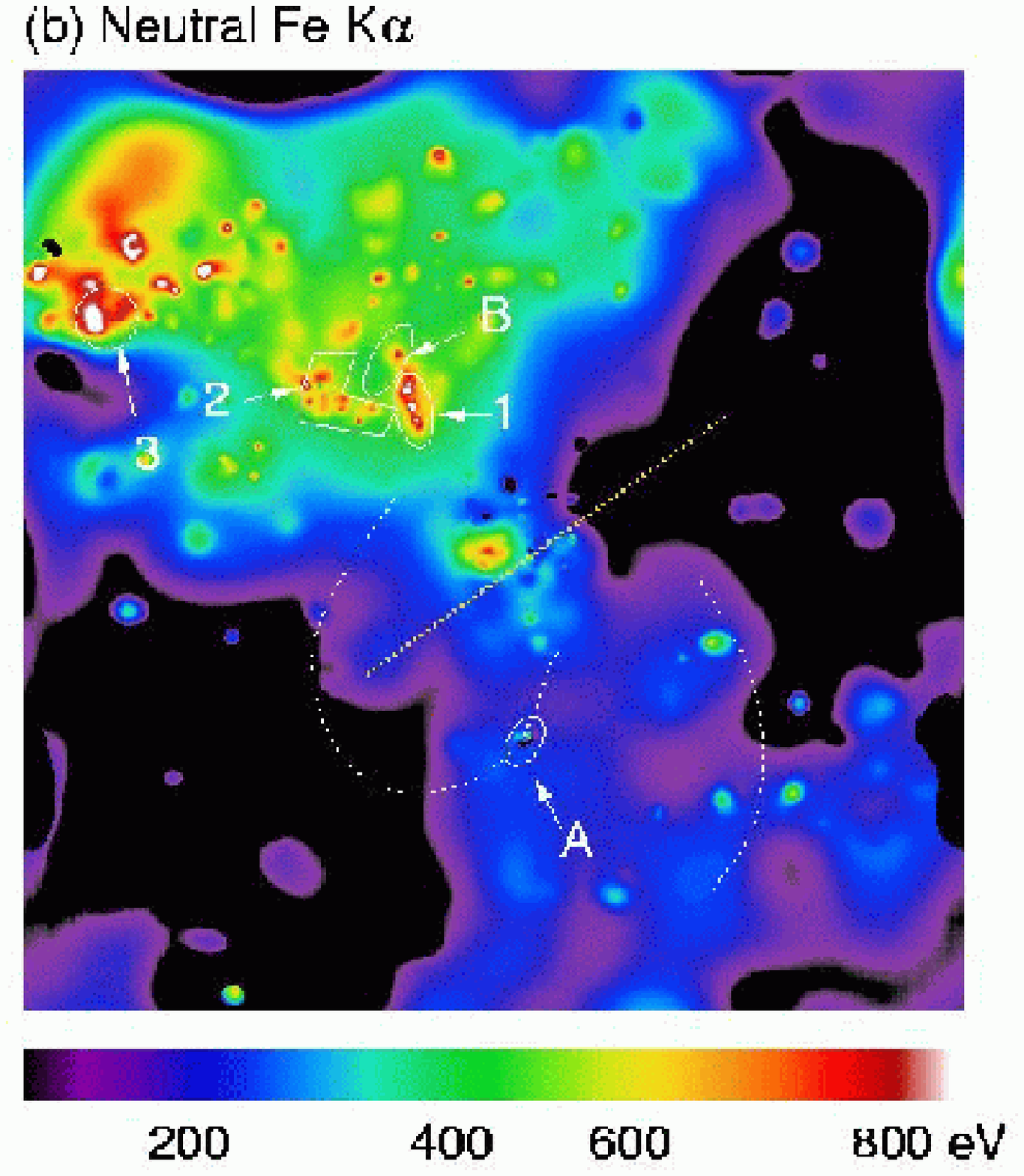}}}
\centerline{{\includegraphics[width=0.4\textwidth]{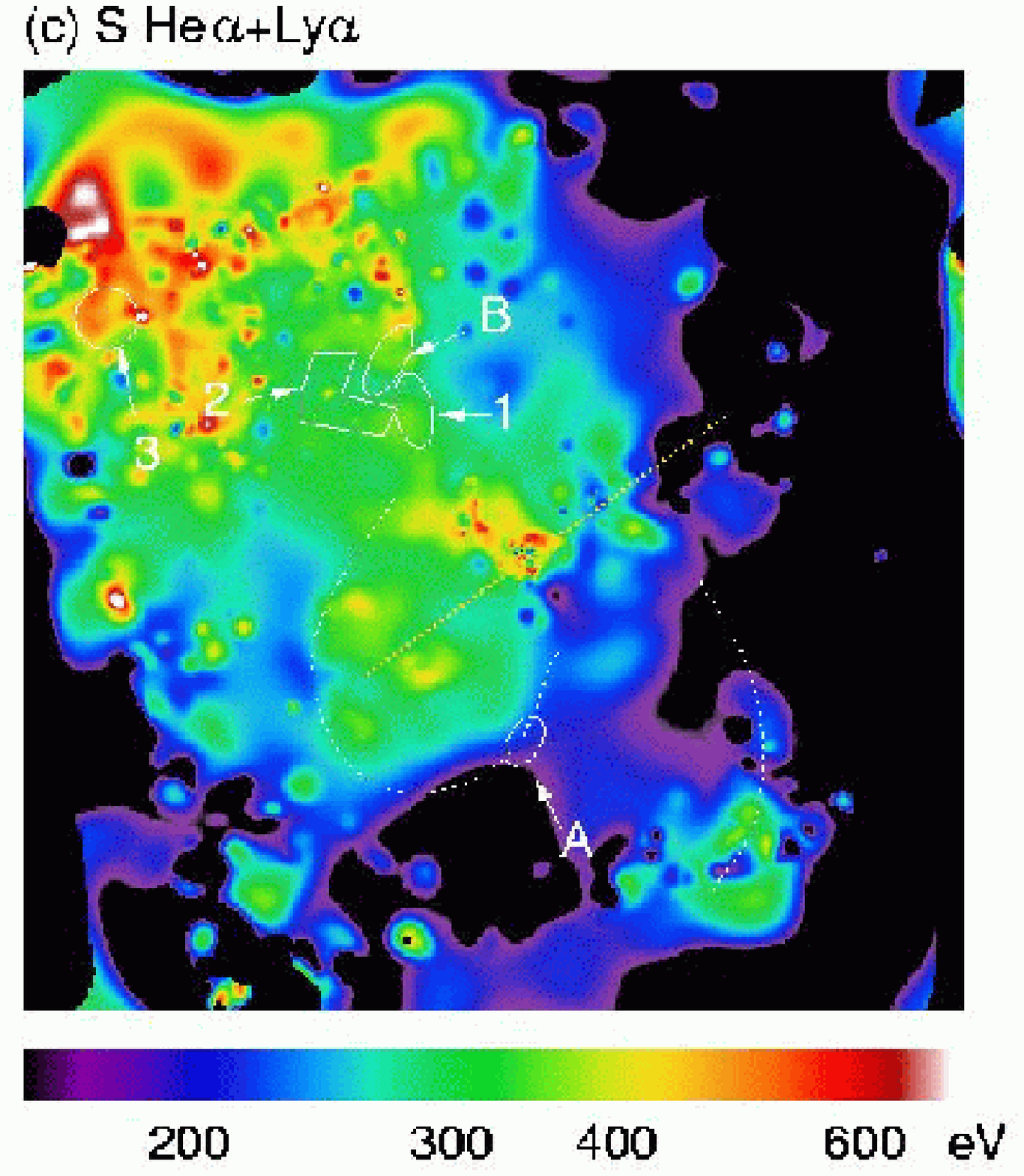}}
{\includegraphics[width=0.4\textwidth]{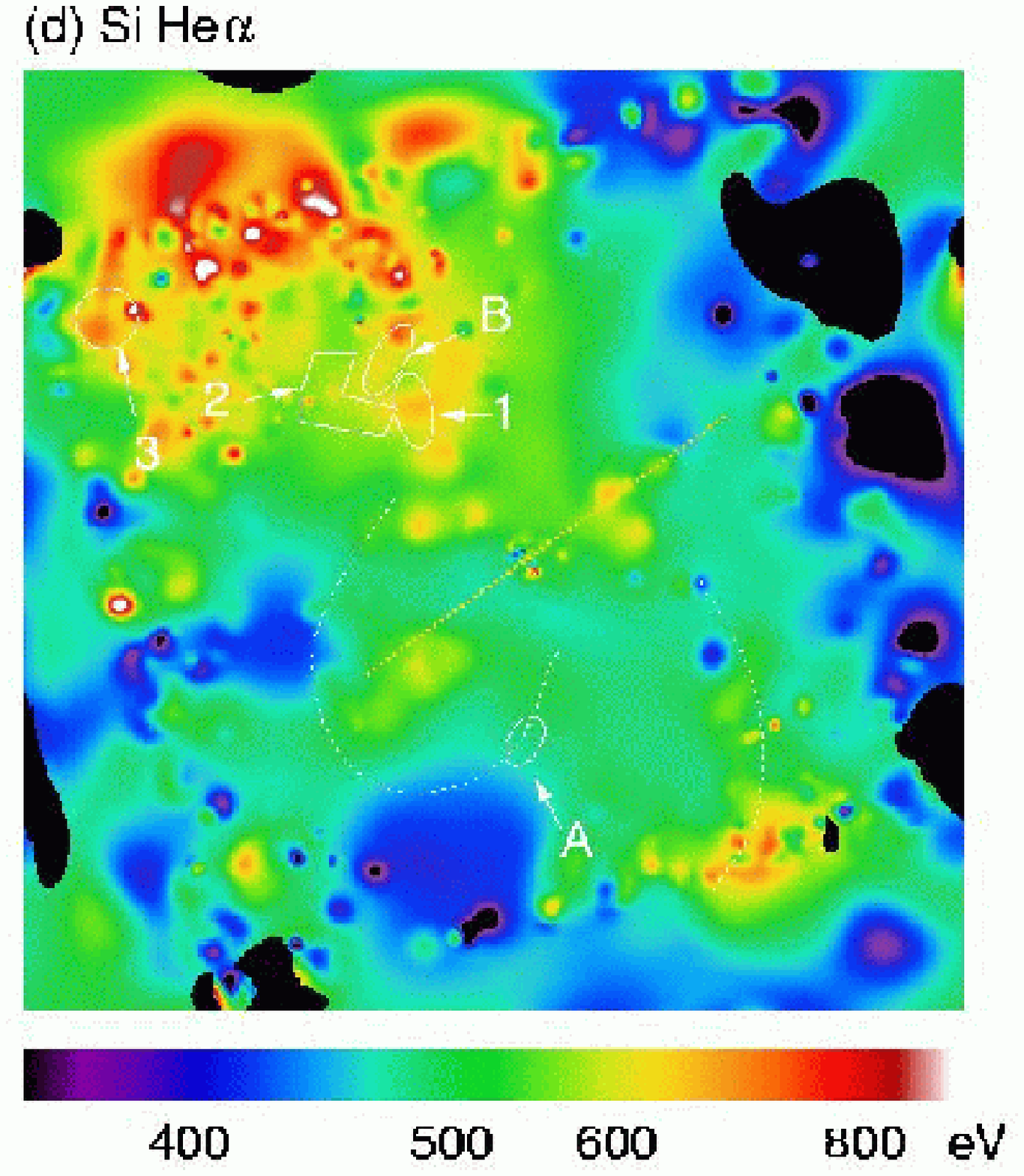}}}
\figcaption[]{False-color EW images of a) Fe He$\alpha$ ($E$ $\sim$ 6.7 
keV), b) ``neutral'' Fe K$_{\alpha}$ ($E$ $\sim$ 6.4 keV), c) S He$\alpha$ 
+ Ly$\alpha$ ($E$ $\sim$ 2.5 keV), and d) Si He$\alpha$ ($E$ $\sim$ 1.8 
keV) lines. The line and continuum images have been adaptively smoothed
prior to calculation of the EW. Zero EW has been assigned where the
estimated underlying continuum is low (typically $\la$5\% of the maximum).
Some regions as identified in Figure~\ref{fig:fig1} are marked for
comparisons.
\label{fig:fig2}}
\end{figure}

In order to examine the overall distributions of the diffuse emission
line features around the Galactic center region, we construct {\it 
equivalent width} (EW) images for the detected atomic emission lines, 
following the method described by Park et al. (2002). After removing 
all detected point sources from the broadband image (see Muno et al. 
[2003a] for the details of the point source detection), subband images 
for the line and continuum bandpasses were extracted for each spectral 
line of interest (Table \ref{tbl:tab2}). These subband images were 
adaptively smoothed to achieve a S/N ratio of 3 $-$ 4. The underlying 
continuum was calculated by logarithmically interpolating between images 
made from the higher and lower energy ``shoulders'' of each broad line. 
The estimated continuum flux was integrated over the selected line width 
and subtracted from the line emission. The continuum-subtracted line 
intensity was then divided by the estimated continuum on a pixel-by-pixel 
basis to generate the EW images for each element. In order to avoid 
noise in the EW maps caused by poor photon statistics near the CCD chip 
boundaries, we have set the EW values to zero where the estimated continuum 
flux is low. We also set the EW to zero where the integrated continuum flux 
is greater than the line flux. We present the EW images of the Fe He$\alpha$ 
($E$ $\sim$ 6.7 keV), the neutral Fe K$_{\alpha}$ ($E$ $\sim$ 6.4 keV), S 
He$\alpha$ + Ly$\alpha$ ($E$ $\sim$ 2.5 keV) and Si He$\alpha$ ($E$ $\sim$ 
1.8 keV) line emission (Figure \ref{fig:fig2}). 
We note that cosmic X-ray background (CXB) and the particle
background (PB), both of which have not been subtracted in the EW
generation, could have contaminated our EW images to some extent. 
We found that the CXB contribution to the observed fluxes is negligible
($\la$1\%; e.g., Muno et al. 2003b). The PB contamination 
appears to be more significant than the CXB and may contribute up to 
$\sim$20\%$-$30\% of the estimated continuum emission where the surface 
brightness is low. The EW estimations are thus affected by 
the PB contamination by up to $\sim$20\%. This embedded contamination 
is not however critical for our purposes of the EW images: i.e., we 
intended to utilize these EW images in order to qualitatively investigate 
overall variation and distribution of the bright X-ray line emission 
over the Galactic center region. We do not use the estimated absolute EW 
values unless the actual regional spectal analysis has been performed.
The adaptive smoothing also introduces some systematic uncertainties
for the small-scale, faint EW features which we do not rely on for
the analysis. The large-scale ($\ga$arcmin), relatively bright EW features
are, on the other hand, consistent with the actual spectral analysis
and are reliable in the context with previous observations (see the 
following discussions and sections). Despite the embedded uncertainties, 
these EW images are therefore useful as first-order information in order 
to investigate the overall variation of the X-ray line emission within the
FOV and to qualitatively discuss some noticeable characteristics.
We also note that the line flux maps as well as the EW maps might also be 
useful to understand the line emission distributions. Since the line
flux maps show the {\it same} overall intensity variations as the EW maps,
we do not explicitly present the line flux maps although our discussion
on the line features is in parts based on the line flux distribution in
addition to the EW variations.

The Fe He$\alpha$ line EW is generally enhanced along the plane. The 
enhancements in the Fe He$\alpha$ line EW of this Galactic plane component 
are marginal (roughly around $\sim$400 eV), which is consistent with the 
{\it ASCA} measurements of the hard Galactic ridge X-ray emission (GRXE) 
component ($E$ $\sim$ 8 keV) \citep{kaneda97}. This marginal EW enhancements 
are also consistent with those from the detected point sources in the 
Galactic center \citep{muno03a}. A fraction of this 
Fe He$\alpha$ EW should thus originate from unresolved point sources. 
The substantial EW variations on arcmin scales, however, suggest that a 
bulk of this hot Fe line emission has a diffuse origin arising from a hot 
($E$ $\sim$ 5 $-$ 10 keV) plasma. Sgr A East is a strong Fe He$\alpha$ 
line source \citep{maeda02}, and is immediately recognized as an extremely 
bright source with an $\sim$1$\farcm$3 diameter in the center of the 
Fe He$\alpha$ EW image (Figure \ref{fig:fig2}a). The estimated Fe 
He$\alpha$ line EW is $\sim$2500 eV $-$ 5000 eV within the central 
$\sim$30$^{\prime\prime}$ diameter region of Sgr A East, which is 
consistent with the results from the spectral analysis by Maeda et al. 
(2002). This strong Fe EW appears to be from hot Fe-rich ejecta heated by 
the reverse shock of the SNR Sgr A East as posited by Maeda et al. (2002).

The neutral Fe EW is largely distributed along the plane, and, in contrast
to the He-like Fe EW, is strongly enhanced in the northeast of Sgr A* 
(Figure \ref{fig:fig2}b). The bright, hard Fe knots in the northeast of 
Sgr A* (regions 1, 2, and 3 in Figure \ref{fig:fig1}) are featureless in 
the Fe He$\alpha$ EW image (Figure \ref{fig:fig2}a), but remarkably bright 
in the neutral Fe EW image (Figure \ref{fig:fig2}b). The regions to the 
northeast of Sgr A* are also bright in the Si and S EW images 
(Figure~\ref{fig:fig2}c and Figure~\ref{fig:fig2}d). The Si and S maps 
represent high ionization states in hot thermal plasmas which are typically 
produced by SNR and/or star-forming activity. The extinction toward the 
hot S and Si emission in the northeast of Sgr A* is consistent with that 
of typical column density toward the Galactic center (see \S\ref{sec:spec},
also Muno et al. 2003b), which indicates the Galactic 
center origin of the hot gas rather than a nearby foreground component. 
It is interesting that these high-ionization-state hot S, Si material and 
the low-ionization-state cold Fe sources are in general coincident in the 
northeast of Sgr A*. On the other hand, albeit that the angular resolution 
and the positional accuracy of the EW images are significantly degraded 
due to the adaptive smoothing of faint diffuse emission, the one-to-one 
correlations between the bright S/Si EW and the neutral Fe EW enhancements 
appear to be weak: i.e., the locations of the strong enhancements of hot 
and cold gas are probably not ``exactly'' coincident. These cold, neutral 
Fe line emitting gas and the hot, ionized S/Si line emitting plasma also 
appear to have Galactic center molecular cloud counterparts (see 
\S\ref{sec:mc}). These features suggest that the high- and 
low-ionization-state line emission might be associated with the 
star-forming/SN activities nearby to the dense molecular cloud complexes 
in the northeast of Sgr A*. Both of S/Si and neutral Fe EWs are weak in 
the southwest of Sgr A*, which appears to be caused by intrinscally less 
activities there (for more discussion, see \S\ref{sec:mc} and 
\S\ref{sec:disc}).

\section{\label{sec:spec} X-Ray Spectra}

We perform a spectral analysis of a few diffuse emission features, 
particularly those with strong neutral Fe EWs. The individual ACIS   
observations comprising our images utilized different roll-angles: i.e.,     
the first three observations have $\sim$270$^{\circ}$ roll-angles and 
the remainder $\sim$80$^{\circ}$. The Fe knots have thus been detected on   
either ACIS-I0 or ACIS-I3 chips depending on the roll-angle. About 
$\sim$85\% of the total exposure was taken with roll-angles of              
$\sim$80$^{\circ}$, however, so the bulk of the photons from these Fe       
knots have been detected on the ACIS-I0 chip.  We thus utilize the 
response matrices appropriate for ACIS-I0.  The Fe knots 1, 2, and 3        
contain $\sim$12000, $\sim$24000, and $\sim$11000 photons, respectively, 
and we bin the data to contain a minimum of 60 $-$ 200 counts per channel 
for the spectral fitting. These emission features are fairly bright and
have well-confined structures, and so we estimate the background 
contamination by extracting spectra from nearby source-free regions
after removing all detected point-like sources from the data.  
For the spectral analysis of our CTI-corrected    
data, we utilize the response matrices appropriate for the spectral         
redistribution of the CCD, as generated by Townsley et al. (2002b).  The 
low energy ($E$ $\la$ 1 keV) quantum efficiency (QE) of the ACIS has 
degraded because of the molecular contamination on the optical blocking 
filter.  Although effects from this QE degradation on our heavily absorbed 
spectra are most likely negligible, we correct this time-dependent QE       
degradation by modifying the ancillary response function (ARF) for each 
extracted spectrum by utilizing the IDL ACISABS software\footnote{For 
the discussion on this instrumental issue, see
http://cxc.harvard.edu/cal/Acis/Cal\_prods/qeDeg/index.html.
The software was developed by George Chartas and is available at
http://www.astro.psu.edu/users/chartas/xcontdir/xcont.html.}.

\begin{figure}[]
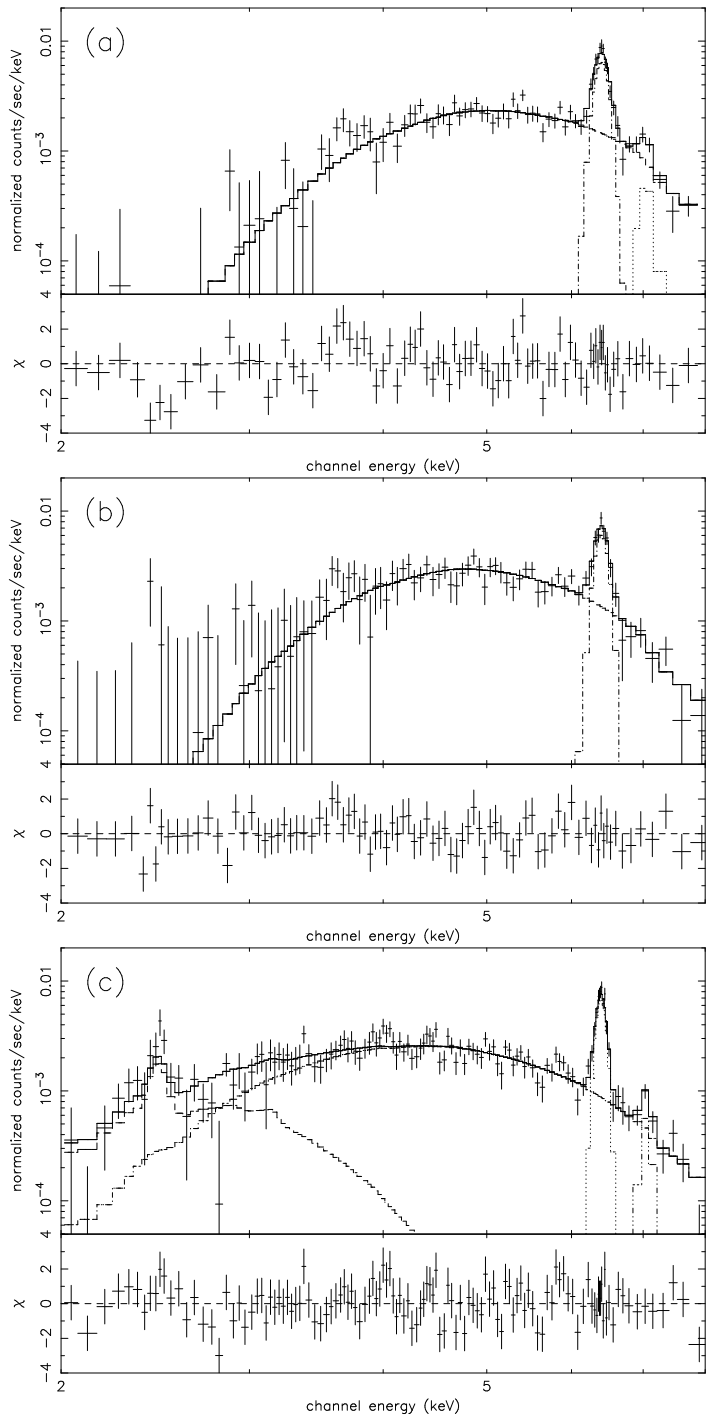

\figurenum{3}
\includegraphics[angle=-90,width=0.5\textwidth]{f3a.ps}\\
\includegraphics[angle=-90,width=0.5\textwidth]{f3b.ps}\\
\includegraphics[angle=-90,width=0.5\textwidth]{f3c.ps}
\figcaption[]{The {\it Chandra}/ACIS spectrum: a) Fe knot 1 , b) Fe 
knot 2, and c) Fe knot 3. In panels (a) and (b), the best-fit power-law
continuum and Gaussian components are presented. In (c), the 
additional soft thermal component is also presented as well as the
hard continuum and Gaussian components.
\label{fig:fig3}}
\end{figure}

The X-ray spectral characteristics of knots 1 and 2 are similar to each     
other: i.e., a strong 6.4 keV Fe emission line with underlying power-law    
continuum of photon index $\Gamma$ $\sim$ 2 $-$ 3 (Figure \ref{fig:fig3}a 
and Figure \ref{fig:fig3}b).  Both spectra are heavily absorbed ($N_H$ 
$\sim$ 3.5 $\times$ 10$^{23}$ cm$^{-2}$) and the photon fluxes are 
insignificant below $E$ $\la$ 4 keV. We thus present the best-fit 
parameters based on spectral fittings in the 4 $-$ 8 keV band (Table 
\ref{tbl:tab3}). The measured EW of the Fe line is $\sim$1 keV for both 
features, which is consistent with that of the Sgr C cloud \citep{mura01a}. 
We note that the knot 1 spectrum may also suggest the presence of an Fe 
K$_{\beta}$ line at $E$ $\sim$ 7 keV (Figure \ref{fig:fig3}a). We have 
added another Gaussian line component in order to fit this weak feature,    
assuming a narrow line width. The best-fit K$_{\beta}$ line center energy 
is 7.02$^{+0.08}_{-0.05}$ keV with an EW of $\sim$200 eV. The measured 
K$_{\beta}$ to K$_{\alpha}$ line flux ratio is 0.14$\pm$0.03. 
This K$_{\beta}$ line feature is not apparent in the knot 2 spectrum, 
at least with the current data. We place an upper limit (2$\sigma$) on 
the K$_{\beta}$ line EW of 120 eV for knot 2.

The Fe knot 3 spectrum shows the same characteristics as the spectrum of    
the other two knots in the hard X-ray band ($E$ $>$ 4 keV): i.e., a strong 
6.4 keV Fe K$_{\alpha}$ line with underlying continuum of $\Gamma$ 
$\sim$ 1.8 (Figure \ref{fig:fig3}c). The measured 6.4 keV Fe line EW is 
$\sim$1.3 keV. This spectrum also appears to show a weak Fe K$_{\beta}$ line 
feature as seen in the knot 1 spectrum. The best-fit line center energy
of this weak feature, the measured EW, and the line flux ratio between
K$_{\beta}$ to K$_{\alpha}$ lines are consistent with those of knot 1. 
The knot 3 spectrum, on the other hand, is less absorbed ($N_H$ $\sim$
1.5 $\times$ 10$^{23}$ cm$^{-2}$) than knots 1 and 2, revealing remarkably 
different features than knots 1 and 2 in the soft band ($E$ $<$ 4 keV; 
Figure \ref{fig:fig3}c). The soft band X-ray spectrum of knot 3 is 
dominated by thermal emission with a strong S He$\alpha$ line. 
For comparison, a single power law model fit is relatively poor 
($\chi^2_{\nu}$ = 1.4) and the inclusion of an additional thermal plasma 
model improves the overall fit significantly based on an F-test. The strong 
S line emission from the soft component of the knot 3 spectrum indicates 
X-ray emission from a thermal plasma having an electron temperature of     
$kT$ $\sim$ 0.3 keV. The metal abundance for the soft component appears
to be high and is best-fitted with a few times solar, although 
unconstrained with the current data (e.g., a 2$\sigma$ lower limit 
is $\sim$0.5 solar). This soft thermal component appears to be 
associated with the Galactic center molecular clouds (\S~\ref{sec:mc}), 
suggesting its location in the Galactic center instead of a nearby 
foreground origin. The spectral fit with a two-component absorption for 
knot 3 indeed indicates a significant foreground column (N$_H$ $\sim$ 1 
$\times$ 10$^{23}$ cm$^{-2}$) even only for the soft thermal component. 
The nearby foreground origin for this soft thermal emission is 
therefore unlikely. The best-fit parameters for the knot 3 
spectrum are presented based on the spectral fitting in the 2 $-$ 8 keV 
band (Table \ref{tbl:tab3}).


\section{\label{sec:mc} Molecular Line Emission}

The spatial distribution and spectral X-ray characteristics revealed 
by the data suggest that some of the diffuse X-ray emission features 
in the Galactic center be associated with the cold, dense component 
of the ISM and/or active star-forming regions.
We thus examine Galactic center molecular cloud maps in order to
investigate correspondences between dense clouds and the detected 
X-ray emission features.  Considering the high molecular densities 
in the Galactic center region, the CS molecule, with its relatively 
high critical density, $n$(H$_2$) $\sim$ 10$^4$ cm$^{-3}$, is an 
appropriate tracer of dense molecular clouds there. The CS line maps 
are, unlike those of CO, expected to be nearly free of strong 
contamination by the Galactic disk and are more representative of the
molecular column density rather than the surface temperatures of clouds
(Tsuboi et al. 1999 and references therein). The CS data also provide 
an angular resolution of $\sim$30$^{\prime\prime}$ (Tsuboi et al. 1999), 
allowing a meaningful cross-comparison with the high resolution {\it 
Chandra} image.  We have thus searched for potential molecular cloud 
counterparts to X-ray features by investigating various velocity 
components of the 49 GHz CS (J = 1$-$0) line emission \citep{tsuboi99}.

\begin{figure}[]
\figurenum{4}
\centerline{{\includegraphics[width=6.5cm]{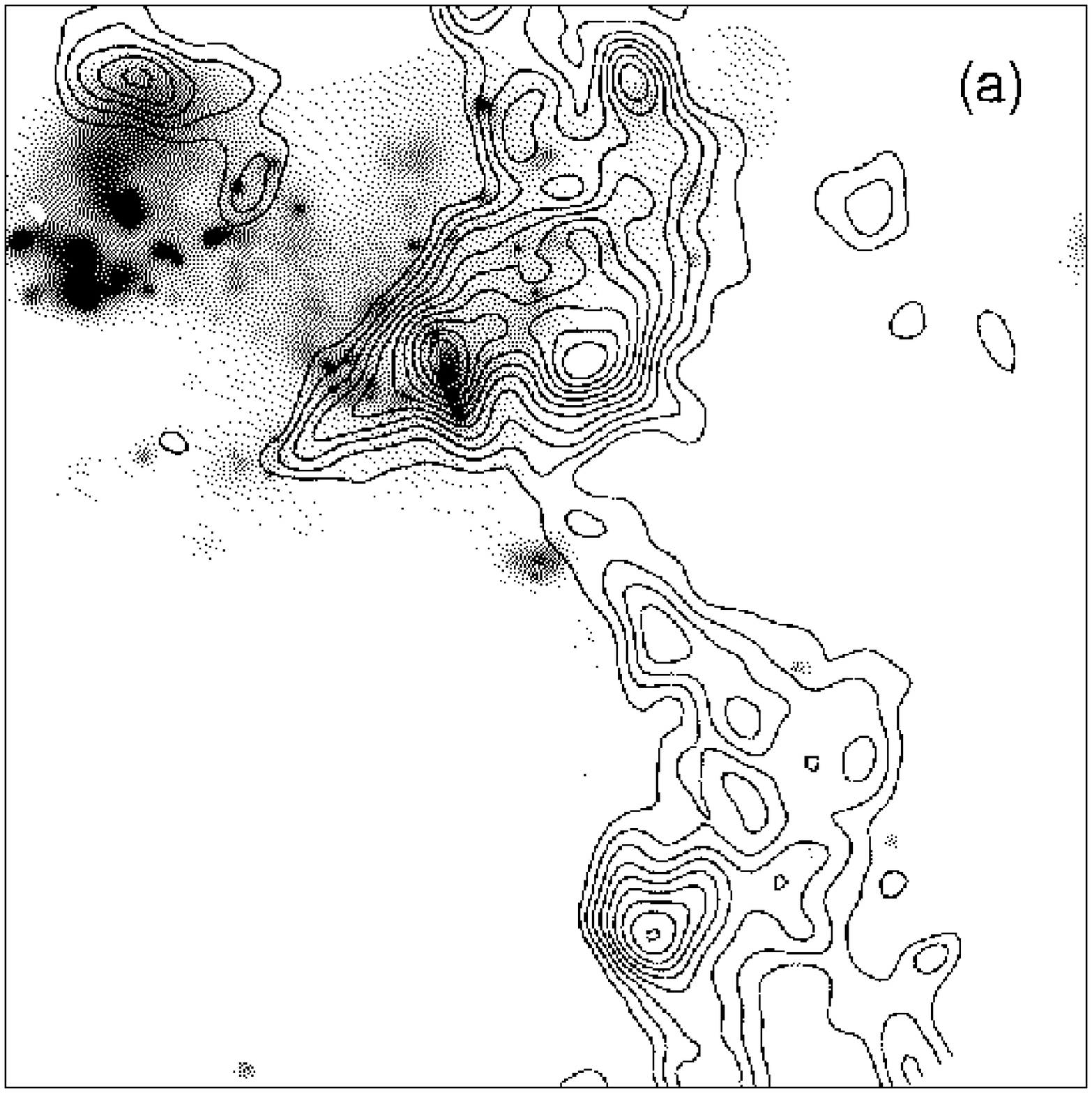}}
{\includegraphics[width=6.5cm]{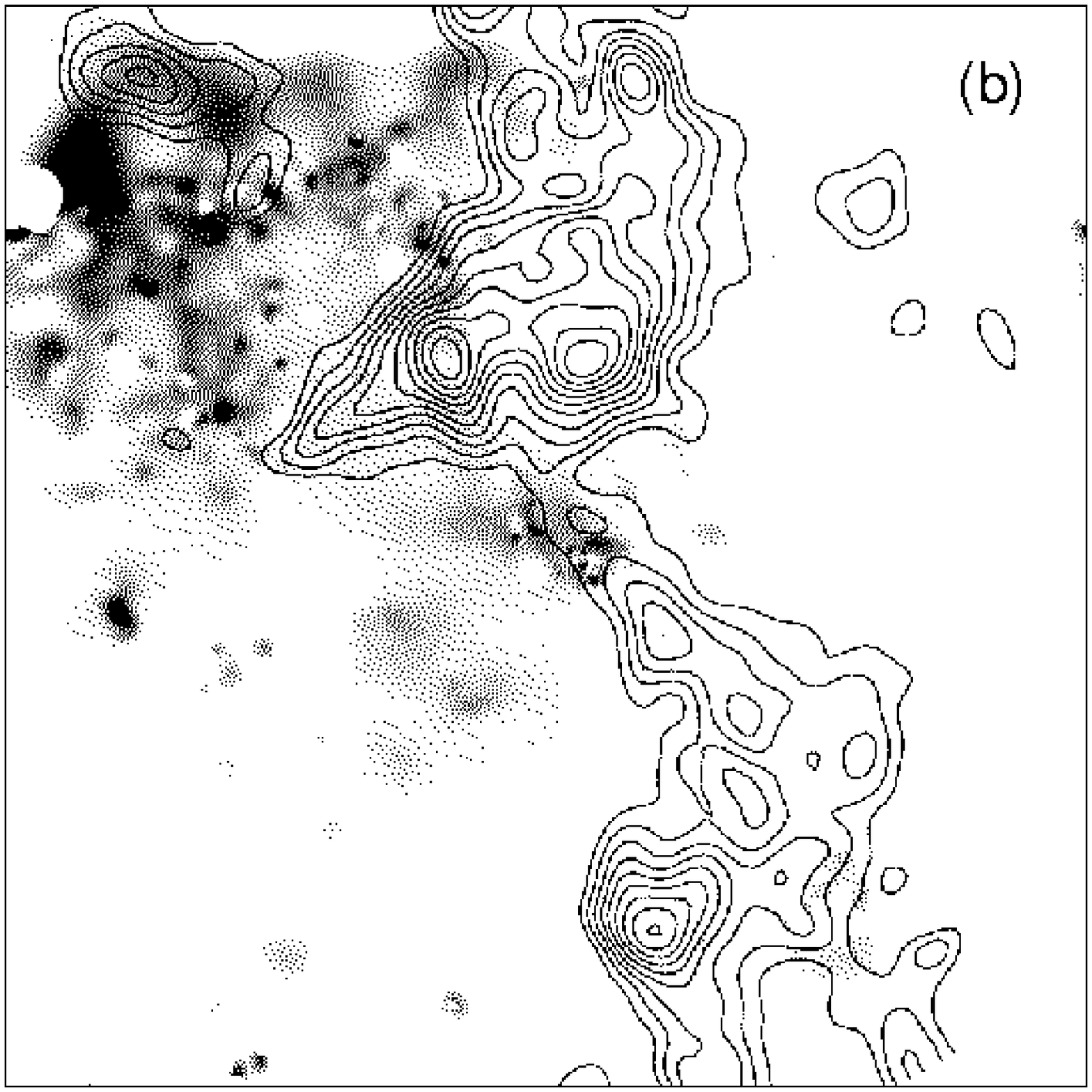}}}
\centerline{{\includegraphics[width=6.5cm]{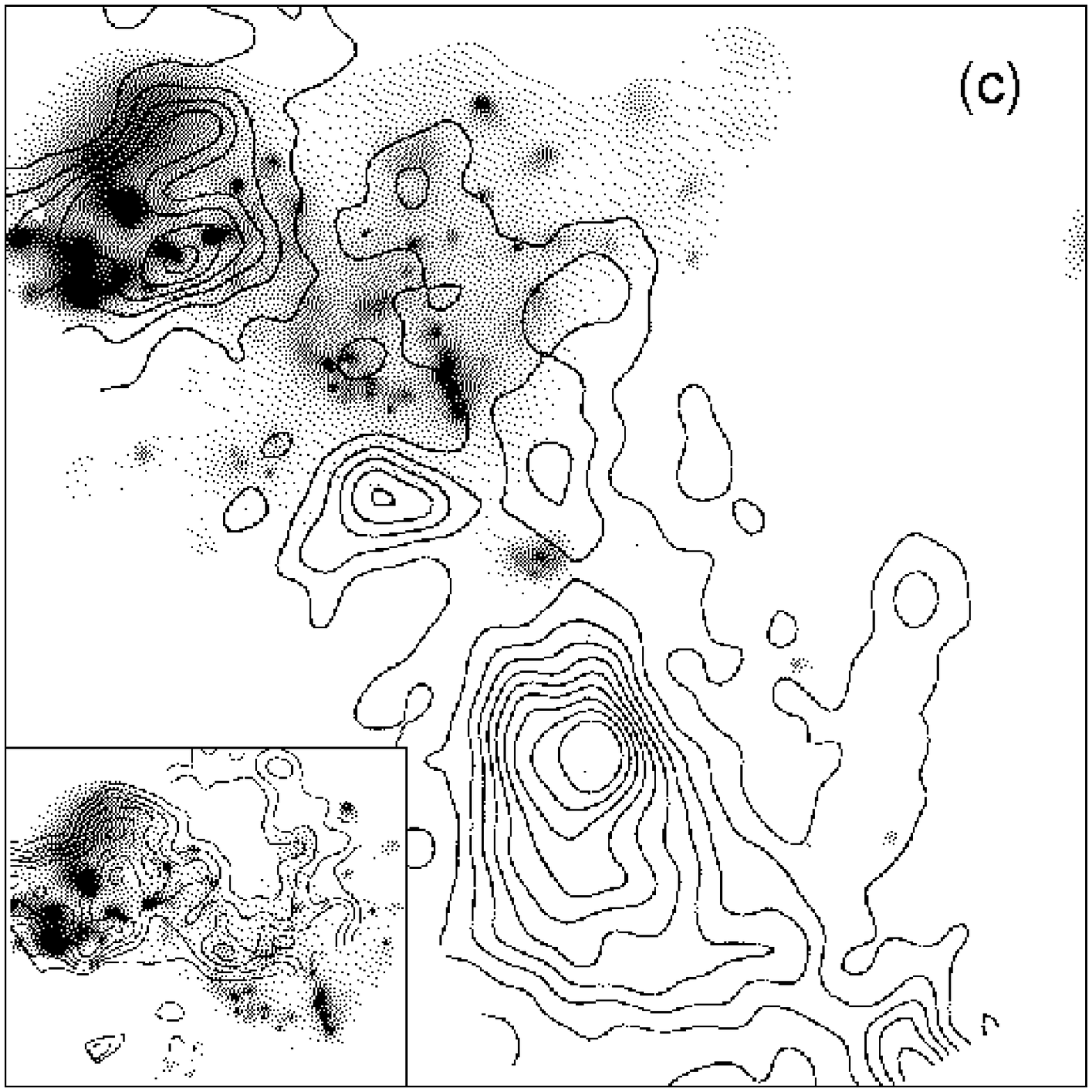}}
{\includegraphics[width=6.5cm]{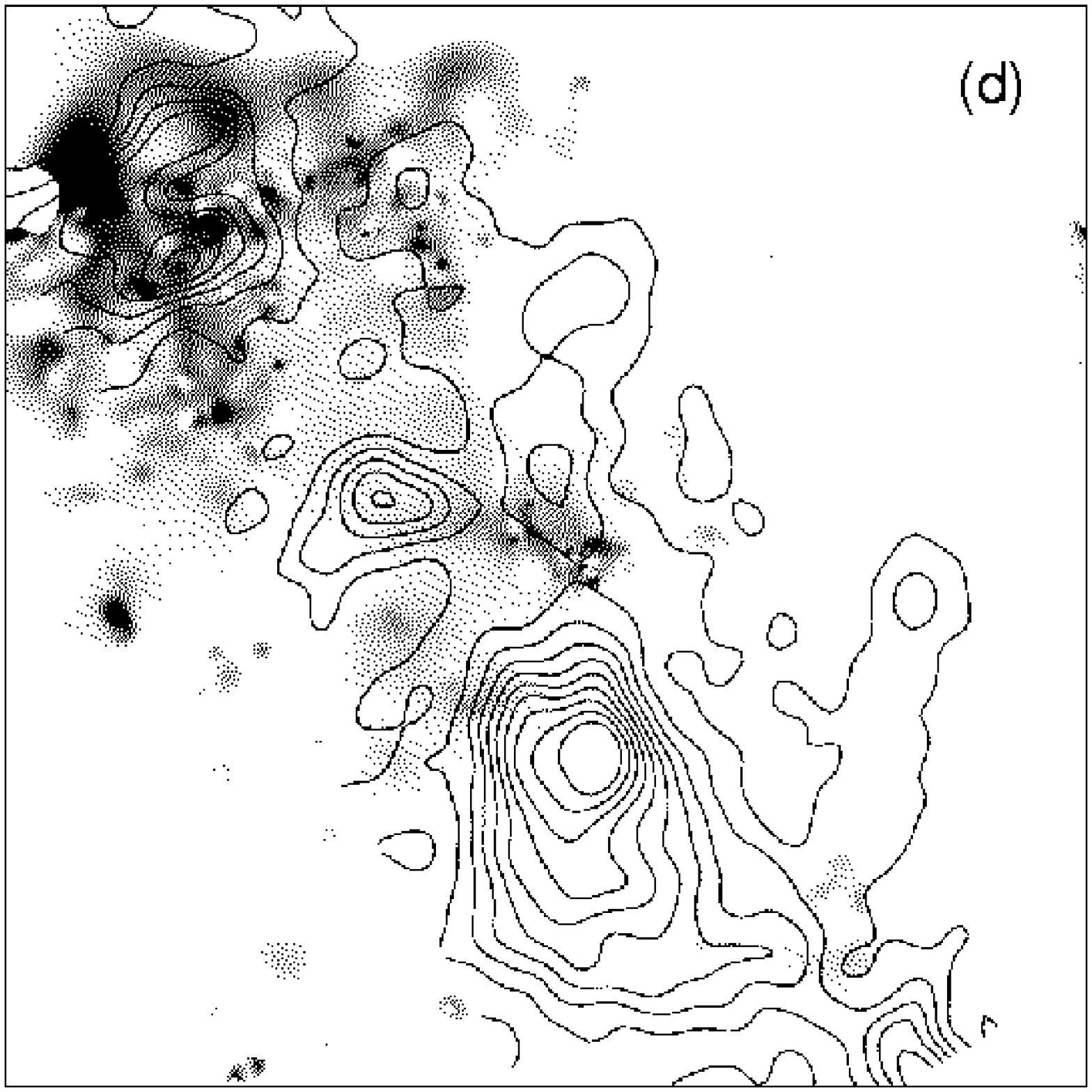}}}
\centerline{{\includegraphics[width=6.5cm]{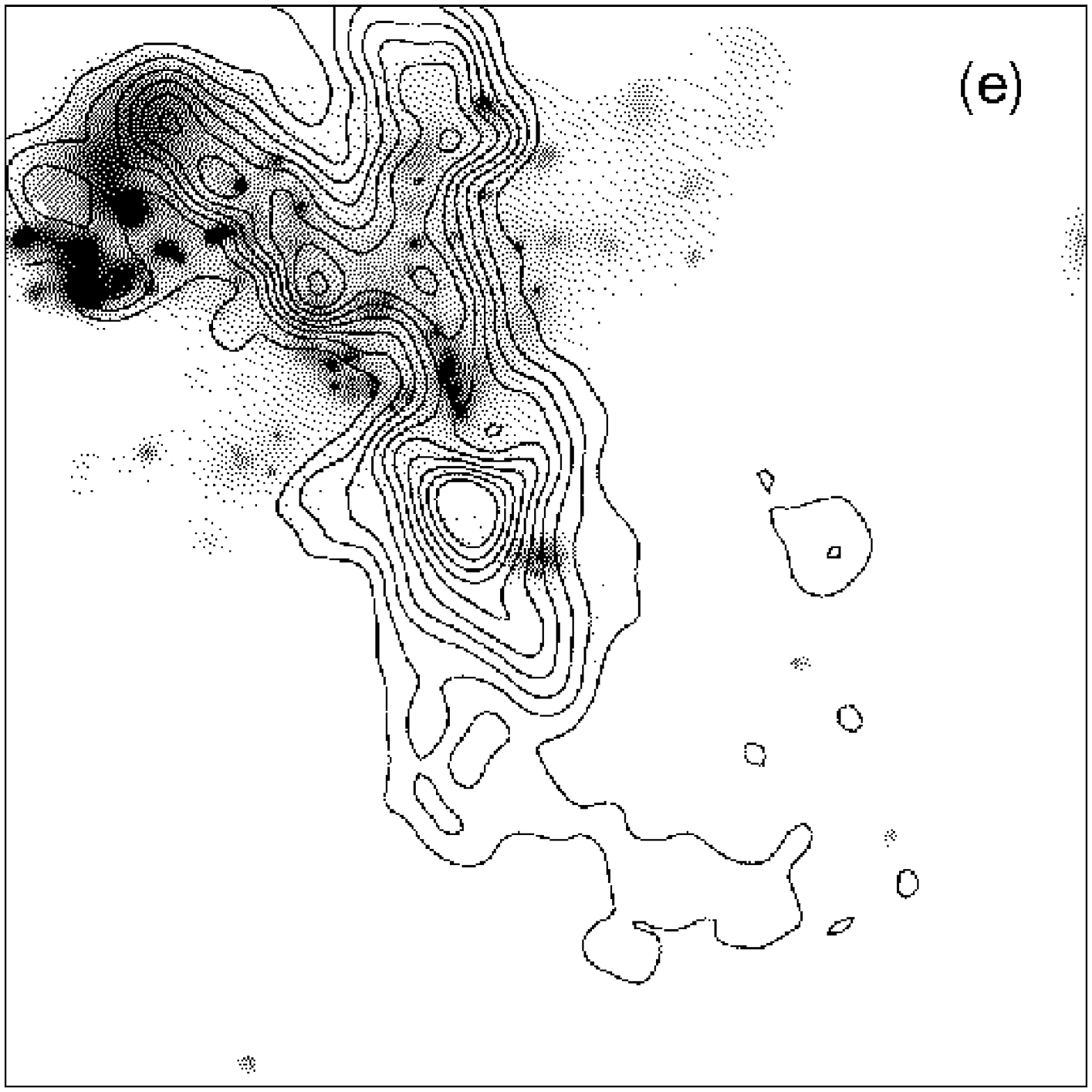}}
{\includegraphics[width=6.5cm]{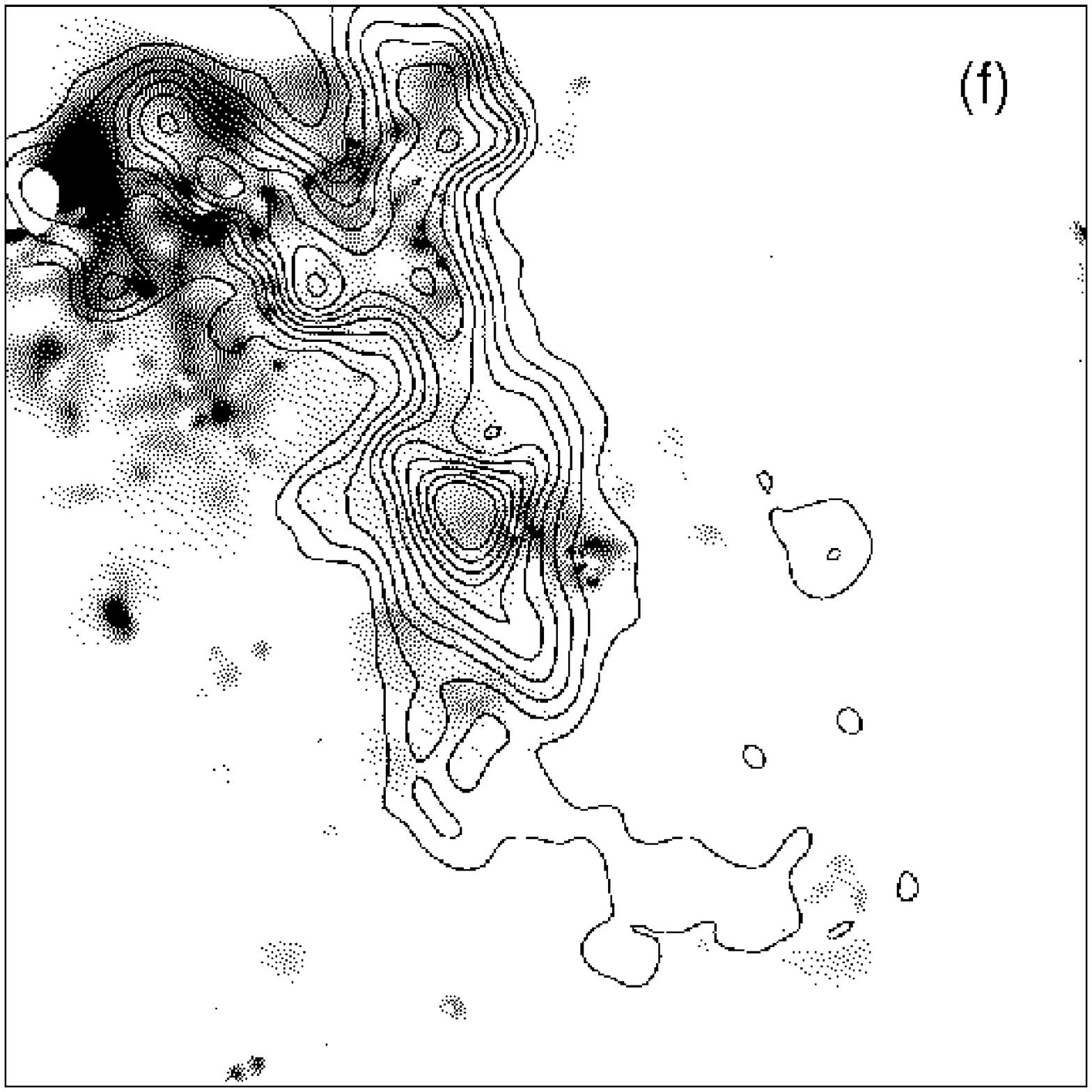}}}
\figcaption[]{Gray-scale EW images of the neutral Fe and the ionized S 
overlaid with CS (J = 1$-$0) contours. a) Fe with CS ($V$ = 
$-$25 to $-$5 km s$^{-1}$), b) S with CS ($V$ = $-$25 to $-$5 km s$^{-1}$),
c) Fe with CS ($V$ = +10 to +30 km s$^{-1}$), d) S with CS ($V$ = +10 to 
+30 km s$^{-1}$), e) Fe with CS ($V$ = +40 to +50 km s$^{-1}$), and f) S
with CS ($V$ = +40 to +50 km s$^{-1}$). The linear scale of the CS contours 
in (a) and (b) ranges 5 to 26 K km s$^{-1}$. The CS contours in (c) and (d)
are 5 to 49 K km s$^{-1}$, and are 5 to 33 K km s$^{-1}$ in (e) and (f).
The lower-left corner inset in panel (c) is the neutral Fe line EW image
overlaid with the 98 GHz CS (J=2$-$1) contours ($V$ = +15 to +45 km 
s$^{-1}$) which ranges 10 to 34.7 K km s$^{-1}$. Although this particular 
CS map has only a partial sky coverage of our {\it Chandra} image (i.e., 
it covers only the northeast region of Sgr A*), the high angular resolution 
($\sim$17$^{\prime\prime}$ HPBW) data reveal the best correlation between 
the CS cloud and the neutral Fe line emission features.
\label{fig:fig4}}
\end{figure}

The overall distribution of dense molecular clouds in the velocity 
range $V$ = $-$25 to $+$50 km s$^{-1}$ is along the Galactic plane, 
and we compare their distributions with our EW images of neutral Fe 
and highly ionized S line emission (Figure \ref{fig:fig4}).  The 
molecular cloud contours in the presented velocity ranges display a 
general correspondence with the X-ray emission-line EW distributions, 
particularly to the northeast of Sgr A*.  The detailed comparisons 
indeed suggest an intimate relationship, but higher spatial resolution 
in the molecular line maps will be required to unambiguously demonstrate 
the nature and geometry of the physical relationship.  The +40 to +50 
km s$^{-1}$ clouds, which, according to Vollmer et al. (2003), are 
located close to, but slightly in front ($\sim$5 pc) of Sgr A*, are 
well correlated spatially with the S and the neutral Fe EW images 
throughout the region to the northeast of Sgr A* (Figure~\ref{fig:fig4}e 
and Figure~\ref{fig:fig4}f).  The most prominent cloud at +10 to +30 
km s$^{-1}$ -- the ``20 km s$^{-1}$ molecular cloud'' located to 
the south of Sgr A* -- is located substantially in front ($\sim$20 
$-$ 50 pc) of Sgr A* \citep{zylka90,vollmer03}, and does not show an
X-ray emission line counterpart.  In this same velocity range, there
is a smaller molecular cloud in the northeast corner of the FOV 
(Figure~\ref{fig:fig4}c and Figure~\ref{fig:fig4}d) which is adjacent 
to the radio Arc and may be interacting with the Arc \citep{tsuboi97}.  
This cloud also appears to be spatially correlated with the most 
prominent emission in the EW images, and with knot 3 in particular. 
This correlation may be best demonstrated by the high angular resolution
98 GHz CS (J=2$-$1) data \citep{tsuboi97} overlaid with the neutral Fe 
line EW image (see the lower-left corner inset in Figure~\ref{fig:fig4}c). 
The line-of-sight locations of the negative 
velocity clouds remains undetermined.  Except for knot 1 (see below),
the molecular emission in the $-$25 to $-$5 km s$^{-1}$ range appears
to be anti-correlated with the S and neutral Fe EW maps in 
the northeast corner of the FOV (Figure~\ref{fig:fig4}a and \ref{fig:fig4}b). 
Some of the bright, hard X-ray filaments have interesting spatial 
correspondences with CS clouds.  Knot 1 falls upon a strong CS line 
peak at negative velocities (Figure~\ref{fig:fig4}a), although it 
also coincides with a prominent ridge in the +50 km s$^{-1}$ cloud 
(Figure~\ref{fig:fig4}e).  Knot 2 is superimposed on the edge of those 
same molecular clouds, so its relationship with those clouds, if any,
is unclear.  And, as mentioned above, Knot 3 has plausible molecular 
counterparts in the velocity ranges +10 to +30 km s$^{-1}$ and 
+40 to +50 km s$^{-1}$ (Figure~\ref{fig:fig4}c and 
Figure~\ref{fig:fig4}e).

Toward the south and the southwest of Sgr A*, the overall broadband 
surface brightness, as well as the Si and S line fluxes, are relatively
low, and little correspondence with molecular gas is evident.  The foreground ``20 km s$^{-1}$'' cloud, as represented by the emission integrated from 
+10 to +30 km $^{-1}$ in Figures \ref{fig:fig4}c and \ref{fig:fig4}d, 
is prominent in this region. This dense molecular cloud complex ($n_H$ 
$\sim$ 2 $\times$ 10$^4$ cm$^{-3}$) implies a significant column density 
($N_H$ $\sim$ 4 $\times$ 10$^{23}$ cm$^{-2}$) and a correspondingly large
visual extinction, $A_V$ $\ga$ 400 mag \citep{zylka90}, so this cloud
should be optically thick in the soft X-ray band.  This possibly accounts
for the observed low intensities in the broadband surface brightness and 
for the soft X-ray line fluxes to the south of Sgr A*.  The relative
lack of detected X-ray point sources toward this 20 km s$^{-1}$ cloud \citep{muno03a} is consistent with a strong absorption effect.  
We also note that the Si, S, and neutral Fe line fluxes and the EWs to 
the southwest of Sgr A* where the negative velocity cloud is present, but
where the 20 km s$^{-1}$ cloud is absent, are weak.  This stands in 
contrast to the strong line emission in Si, S, and neutral Fe toward 
dense clouds in the northeast portion of the FOV.  The low line fluxes 
in this southwest region cannot be caused by absorption in the 20 km s$^{-1}$ cloud simply because of the absence of this cloud in this direction. 
It is also unlikely to be caused by absorption in the negative velocity cloud 
because there is no evidence for substantial obscuration of the X-ray point 
sources in this southwest region toward the negative velocity cloud 
\citep{muno03a}.  Thus, the weak X-ray line features to the southwest of 
Sgr A*, where the 20 km s$^{-1}$ cloud is absent, appear to be intrinsic: 
that is, this region is lack of supernova activity, or perhaps
of energetic star-forming activity compared to the northeast regions.

Based on the spectral analysis presented in \S\ref{sec:spec}, we
realize that the characteristics of the knot 3 spectrum 
are consistent with those of knot 1, except for the low absorbing
column density which reveals the soft thermal component. The column 
density toward knot 3 ($N_H$ $\sim$ 1.5 $\times$ 10$^{23}$ cm$^{-2}$) 
is consistent with typical columns to the Galactic center \citep{sakano98}. 
Knot 1 (and 2), on the other hand, has significantly higher column 
($N_H$ $\sim$ 3.5 $\times$ 10$^{23}$ cm$^{-2}$). We speculate that the 
larger column for knot 1 (and 2) might be attributable to the negative 
velocity molecular cloud in this direction.  Consistent with this 
hypothesis is the fact that SgrA East, which has the same column as 
knot 3 \citep{maeda02}, is not coincident with emission in the $-$25 
to $-$5 km s$^{-1}$ CS map (Figure \ref{fig:fig4}).  The apparent angular 
size of the $-$25 to $-$5 km s$^{-1}$ cloud to the northeast of Sgr A* is 
$\sim$6$'$, so the physical extent is $\sim$15 pc.  If the bulk 
of the excess column toward knot 1 ($N_H$ $\sim$ 2 $\times$ 10$^{23}$ 
cm$^{-2}$) is attributed to this cloud, a cloud density of $n$ $\sim$ 4 
$\times$ 10$^{3}$ cm$^{-3}$ is implied.  Assuming a simple spherical 
geometry for the cloud, the total mass of the cloud is $M$ $\sim$ 2 
$\times$ 10$^5$ $M_{\odot}$.  Albeit crude estimates, the derived density 
and mass of the cloud are not unreasonable compared with other giant 
molecular cloud complexes in the Galactic center.  Based on these X-ray
absorption features, we therefore propose that this cloud is in the 
``foreground'' of Fe knot 1 (and 2), although some portions appear to 
be associated with knot 1 itself.  The soft X-ray emission to the 
northeast of Sgr A*, as presented by the S EW images (Figure 
\ref{fig:fig4}b), is apparently anti-correlated with the negative velocity 
CS emission, which can be accounted for by the presence of this absorbing 
foreground cloud.

\section{\label{sec:disc} DISCUSSION}

\subsection{\label{subsec:nonthermal} Neutral Iron Line Emission}

The bright, hard X-ray filaments to the northeast of Sgr A* stand 
out for having strong neutral Fe emission. The spectral 
characteristics of Knot 1, for example, are consistent with those 
of the Sgr B2 and Sgr C clouds, both of which have been plausibly 
described by an XRN model illuminated by past activity of Sgr A* \citep{mura00,mura01a,mura01b}.  Knot 1 has a cometary morphology, 
with the ``head'' in the north and extended south by $\sim$1$'$, 
along a minor ridge of the negative velocity CS cloud.  The spectral 
features of the strong 6.4 keV Fe K$_{\alpha}$ line, a weak 7.02 keV 
Fe K$_{\beta}$ line, and the underlying continuum with $\Gamma$ $\sim$ 
2 are typical for the proposed XRN model \citep{koyama96,mura00}.  
While the Fe K$_{\alpha}$ EW of $\ga$1 keV is weaker than that from 
the Sgr B2 XRN \citep{mura00}, it nonetheless suggests that the Fe line 
emission is dominated by irradiation from external sources (Sunyaev 
\& Churazov 1998 and references therein).  Since the Thomson 
cross-section is dependent on the scattering angle of the incident 
photon, while the fluorescent line emission is isotropic, the observed 
EW -- the ratio of fluorescent line intensity to the scattered continuum
-- is inversely proportional to 1 + cos$^2$$\theta$, where $\theta$ is 
the angle subtended at the scatterer between the observer and the 
illuminating source.  
If we assume that Sgr A* was the primary external illuminating source 
for knot 1, the lower Fe EW for the knot 1 spectrum, relative to that
of Sgr B2, therefore suggests that $\theta$ is near 0 or $\pi$ and thus
that the primary reflecting cloud toward knot 1 is located in front of 
or behind the plane between Sgr A* and Sgr B2.  (The EW may as well 
depend on other cloud parameters, such as the optical depth, abundances, 
and the ionization state.  Extensive studies of the determinants of the
EW characteristics may thus require detailed XRN simulations, which 
are beyond the scope of the current work.)  The Fe K$_{\beta}$ to 
K$_{\alpha}$ line flux ratio, $\sim$0.15, is consistent with the 
laboratory prediction for the XRN model \citep{mura01b}. 
The sharp absorption edge at $E$ $\sim$ 7.1 keV 
is also in good agreement with previously reported XRN interpretations 
\citep{mura01a,mura01b}. The strong neutral Fe line emission from Knot 1 
is thus well described by the XRN model. Knot 2 shares the overall 
spectral characteristics of knot 1.  However, the molecular counterpart 
is weak or absent for knot 2, and the apparent absence of the Fe 
K$_{\beta}$ line in the knot 2 spectrum is problematical for the XRN
model. The identification of knot 2 as an XRN might thus be inconclusive.

The knot 3 spectrum shows similar features to the knot 1 spectrum in the
hard band ($E$ $>$ 4 keV), but reveals a substantial contribution from a 
thermal component in the soft band ($E$ $<$ 4 keV). The implied electron
temperature ($kT$ $\sim$ 0.3 keV) and likely high abundances might be
suggestive of emission from SN ejecta heated by a blast wave reverse shock. 
This suggested contribution from SNR(s) to the observed knot 3 spectrum
is not surprising because there are dense molecular cloud complexes in 
these regions of the Galactic center, which may contain active 
star-forming regions. In fact, the CS maps indicate associations of 
dense clouds with the X-ray emitting material at the position of knot 3 
(Figures~\ref{fig:fig4}c,\ref{fig:fig4}d,\ref{fig:fig4}e, and 
\ref{fig:fig4}f). 

Based on the XRN model, a past state of dramatically enhanced activity of 
Sgr A*, with an X-ray luminosity of $L_X$ $\ga$ 10$^{39}$ ergs s$^{-1}$, 
has been proposed as the primary external illuminating source for the Sgr 
B2 and Sgr C clouds \citep{mura00,mura01b}. The overall large-scale 
geometry of the neutral Fe EW enhancements to the northeast of Sgr A* 
is consistent with the Fe emission lying predominantly within a conical
region with an apex toward Sgr A* (Figure \ref{fig:fig2}b).  
We thus consider that Sgr A* might have also been the external irradiating 
source for the detected neutral Fe knots. In the reflection models, the 
luminosity of the irradiating external source can be derived in terms of 
the observed 6.4 keV photon flux, the distance to the reflecting cloud 
from the source, and the apparent angular size of the cloud as viewed 
from the source position \citep{sunyaev98,mura00}. Assuming a solar 
abundance for Fe and a Thomson scattering depth $\tau$$_T$ $\sim$ 0.1 
($N_H$ $\sim$ 10$^{23}$ cm$^{-2}$), we derive an irradiating luminosity 
from Sgr A*, $L_X$ $\sim$ 10$^{35}$ ergs s$^{-1}$ for knot 1 and $L_X$ 
$\sim$ 10$^{36}$ ergs s$^{-1}$ for knot 3. These luminosities are $\sim$2 
$-$ 3 orders of magnitude higher than that of the quiescent state flux of 
Sgr A* \citep{baganoff01}, but only $<$0.1\% of those which might have 
produced the XRNe in Sgr B2 and Sgr C clouds. The X-ray luminosity of 
Sgr A* from the recently detected $\sim$3 hr flaring event ($L_X$ $\sim$ 
10$^{35}$ ergs s$^{-1}$; Baganoff et al. [2001]) is comparable to the derived 
irradiating luminosity. Considering the physical sizes of these Fe knots 
($\sim$2 pc for the angular sizes of $\sim$1$'$ at $d$ = 8 kpc), however, 
the duration of the irradiating event must have been a few years or longer 
instead of on the order of $\sim$hours. The inferred Sgr A* activity, if 
it was the irradiating source, should thus have been persistent much 
longer than the observed short flares. 

The inferred irradiating luminosities might have also been generated by 
Galactic transient sources such as low mass X-ray binaries. There are 
likely a number of transient sources in active regions such as the 
Galactic center. In fact, several transient X-ray sources have been 
detected near the Galactic center with X-ray luminosities of $L_X$ 
$\sim$ 10$^{35}$ $-$ 10$^{38}$ ergs s$^{-1}$ (e.g., Branduardi et al. 
1976; Maeda et al. 1996; Pavlinsky et al. 1994). Although durations of 
X-ray transient outbursts can be as long as a decade, typical burst 
periods are relatively short on the order of months. (e.g., Wijnands 
2001). The nearby X-ray transient sources may thus not be completely 
ruled out as contributors to the observed neutral Fe line emission, but 
unlikely. In terms of continuous irradiation sources, it is noteworthy 
that the derived irradiating fluxes are consistent with the X-ray luminosity 
of Sgr A East \citep{maeda02}. The sky position of Sgr A East is between 
Sgr A* and the Fe knots, and so the steady emission from Sgr A East might 
have also contributed the reflected neutral Fe line emission at the 
positions of the observed Fe knots. The measured temperatures for Sgr A 
East, $kT$ $\sim$ 2 keV with {\it Chandra} \citep{maeda02} and $kT$ $\sim$ 
4 keV with {\it XMM-Newton} \citep{sakano02}, however, appear to be too 
soft to produce a strong Fe line at $E$ = 6.4 keV.  Moreover, if Sgr A 
East is a steady illuminating source capable of producing the observed XRN 
features, a strong 6.4 keV Fe line feature would also have been expected from 
the nearby neutral ISM, for example, the circumnuclear disk of Sgr A* 
(e.g., Latkoski et al. 1999 and references therein), which is apparently
not the case.  Steady emission from Sgr A East is thus ruled out as 
a primary irradiating source.

Alternatively, the characteristic 6.4 keV Fe line emission may originate 
from unresolved SN ejecta.  Fast nonthermal particles accelerated by the
Fermi mechanism in collisionless magnetohydrodynamic shocks diffuse 
through cold, metal-rich ejecta knots, and produce low-ionization-state 
K$_{\alpha}$ line X-ray emission \citep{bykov02}. This is most effective 
when the ejecta knot is propagating through dense clouds ($n_H$ $\sim$ 
1000 cm$^{-3}$) \citep{bykov02}. The model predicts an underlying, 
non-thermal bremsstrahlung continuum which is harder ($\Gamma$ $\sim$ 1.5)
than that expected from the XRN model ($\Gamma$ $\sim$ 2). X-ray line 
broadenings caused by the fast-moving unresolved SN ejecta are also 
expected.  The presence of the soft thermal component, particularly in 
knot 3, could be apparent evidence for such a contribution from 
unresolved SN ejecta. Although the observed continuum of knot 3 appears 
to be softer than the typical prediction from the SN ejecta model, the 
large uncertainties on the best-fit photon index of the observed 
underlying continuum still make the ejecta model a plausible candidate
to describe the spectum. 
Although a soft thermal component is not detected from knot 1 and knot 
2, the neutral Fe line characteristics are nearly identical for all of 
these knots. The absence of soft thermal component from knot 1 and knot 
2 may thus be merely due to absorption, and then the supernova ejecta 
model cannot be ruled out as a source of knot 1 and knot 2 as well. 
The estimated Fe line luminosity ($L_X$ $\sim$ 10$^{33}$ ergs s$^{-1}$) 
can be produced by fast-moving ejecta from a single SN explosion 
interacting with a dense molecular cloud \citep{bykov02}.  We note that 
large-scale correlations between the neutral Fe line and the SiO 
(J=1$-$0) emission in the Galactic center regions have been observed 
\citep{martin00}. This gas phase SiO emission due to grain processing 
by strong shocks, being spatially coincident with the 6.4 keV Fe line 
features, may support the SN ejecta origin of the neutral Fe line 
emission in the Galactic center. 

One way to distinguish between the XRN and SN ejecta models is to make 
conclusive measurements of the Fe line broadenings. The detector position 
of knot 3 is close to the CCD readout (rows $<$ 256), so that the energy 
resolution is achieved down to the instrumental limit ($\sim$150 eV FWHM) 
\citep{garmire03}. The measured line width of the 6.4 keV line is then 
$\sigma$ = 10$^{+10}_{-5}$ eV for the fitted Gaussian line profile. This 
is only a weak detection of line width, yet suggests that the Fe line 
width might have been resolved, thus indicating a line broadening in 
this region. If confirmed (by follow-up observations), this line 
broadening would provide support for the contribution from unresolved 
SN ejecta to the observed neutral Fe line emission. Observations with 
instruments of good spectral resolution such as {\it Astro-E2} should 
be helpful for this purpose.  
Better constraints on the photon index of the underlying continuum 
may also be helpful in order to understand the detailed nature of these 
features. Follow-up deep observations with {\it XMM-Newton}, by utilizing
the high sensitivity compared to the {\it Chandra}/ACIS, would be useful 
for this purpose. The SN ejecta model also predicts strong infrared 
emission from the radiative shocks that accompany ejecta fragments
\citep{bykov02}, and so infrared line observations would help to 
discriminate models as well. Although the SN ejecta contribution to
the observed neutral Fe line emission appears to be plausible, we
note a caveat that non-thermal radio emission features have not been 
reported in the northeast regions of Sgr A* within the ACIS FOV. 
This absence of the non-thermal radio features is thus admittedly a 
difficult problem in terms of the SNR interpretation.

\subsection{\label{subsec:thermal} Thermal X-Ray Emission}

The presence of highly ionized elemental species, Si, S, and Fe indicates
that thermal diffuse X-ray emission, as well as the nonthermal component,
also prevails in the Galactic center. These emission lines from the He-like
species over the wide range of photon energies ($E$ $\sim$ 2 keV for 
Si/S and $E$ $\sim$ 7 keV for Fe) confirm the multiple-temperature origin
for the diffuse thermal emission in the Galactic center/plane as indicated 
by the {\it ASCA} data (e.g., Kaneda et al. 1997). The thermal components 
are, with some exceptions of bright features near Sgr A*, relatively faint 
and large-scale features.  The direct spectral analysis of this
diffuse thermal emission thus requires substantially more careful
treatment than the bright, well-confined, nonthermal neutral Fe line
features.  We will present the extensive spectral analysis of the
Galactic center thermal diffuse X-ray emission in subsequent papers
(Muno et al. 2003b; Morris et al. 2003, in preparation).
We here qualitatively discuss the overall characteristics of the thermal 
emission based primarily on the image analysis.

The Si and S line EWs trace enhancements of the soft component of
the thermal emission, which is enhanced to the northeast of
Sgr A* (Figure \ref{fig:fig2}c and Figure \ref{fig:fig2}d).  These
regions are close to the radio Arc and the Arches cluster, where 
the formation of massive stars has taken place in at least two episodes
over the past 5 million years (Figer et al. 1999; Yusef-Zadeh et al. 
2002).  Massive Galactic center molecular clouds are present in these
regions (c.f., \S~\ref{sec:mc}), so the observed Si and S line emission 
most likely arises from the hot ISM heated by powerful stellar 
winds and/or supernova blast waves.  As we have discussed in previous 
sections, the overall broadband X-ray surface brightness and the 
S/Si EWs are low in the southwest regions, which is uncorrelated with 
the absorption effect.  This suggests that the active star forming 
processes are present {\it only} in the northeast of Sgr A* and that
such activity is not manifested in the southwest.  Knot 3 provides a 
sample spectrum of this soft thermal component.  As presented in the 
previous section, this thermal spectrum shows characteristics typical
of hot plasma produced by SNRs.  The {\it ASCA} observations of the 
GRXE have successfully demonstrated that the soft component emission 
in the plane can be furnished by SNRs with a SN rate of $\sim$one 
per century \citep{kaneda97}.  For the case of knot 3, the angular 
size of the emission region ($\sim$1$'$) corresponds to $\sim$2 pc 
at a distance of 8 kpc, which is typical for young Galactic SNRs, 
but might also be of a relatively older age in a dense environment 
such as the Galactic center. 

The hard thermal component is representd by the He-like Fe line EW
(Figure~\ref{fig:fig2}a).  Except for the brightest central feature
of Sgr A East, the highly ionized Fe line emission is primarily 
distributed along the plane.  In contrast to the other EWs in 
Figure~\ref{fig:fig2}, this hot Fe emission shows no particular 
enhancements in the northeast of Sgr A*.  The {\it ASCA} data suggested 
that this hot component might not be adequately described by the SNRs, 
because such a hot plasma may not be confined in the plane by the 
Galactic gravity and/or by typical interstellar pressure \citep{koyama86a}. 
In order to maintain this hot plasma in the plane, a high SN rate and a
low ionization state of the plasma are required \citep{koyama86b,yamauchi95}. 
The electron temperature of the plasma required to produce the observed 
strong 6.7 keV Fe line emission ($kT$ = 5 $-$ 10 keV) is also unusually 
high for typical SNRs, although we do detect the 6.7 keV Fe line 
from metal-rich ejecta in young SNRs (e.g., Hughes et al. 2000; Lewis et al. 
2003). This hot gas with $kT$ $\sim$ 5$-$10 keV might rather have been 
produced when cooler plasma is confined and compressed by strong Galactic 
magnetic fields \citep{makishima97}. 
The required magnetic field strength to achieve this is $B$ = 20$-$30 
$\mu$G.  The interstellar magnetic field in the Galactic center ($B$ 
$\sim$1 mG, e.g., Yusef-Zadeh \& Morris 1987; Lang et al. 1999) 
is sufficiently high for this model. This magnetic confinement model
might not require recent massive star-forming activity or SN explosions
in order to produce the observed hot plasma. The relatively uniform
distribution of the He-like Fe EW along the plane may also indicate
this hot plasma's association with the strong magnetic field in the 
Galactic center along the plane rather than with star-forming regions.

\section{\label{sec:summary} SUMMARY AND CONCLUSIONS}

Deep {\it Chandra}/ACIS observations of the Galactic center reveal
a complex spatial and spectral structure of diffuse X-ray emission with
significant photon statistics.  The thermal (e.g., hot Si and S) and 
nonthermal (e.g., neutral Fe) X-ray line emission is generally enhanced 
in a region lying to the northeast of Sgr A*, a region in which massive
star-formation has been active over the past 5 million years (e.g., the 
Arches cluster), and in which dense Galactic center molecular clouds
are currently present.  The highly ionized line emission from Si and S 
in these regions is therefore most likely associated with the hot ISM 
heated by the shocks from SNe and/or massive stellar winds, while shocks interacting with dense, cold ISM may produce emission in the neutral
Fe line.  Curiously, we detect little or no enhancement of the emission 
lines to the south and southwest of Sgr A*.  This weak line emission to 
the south is apparently the result of an absence of activity there, in
contrast to the northeast regions, and it is owed in part to the substantial
absorption by the foreground 20 km s$^{-1}$ cloud.  In contrast, 
the highly ionized He-like Fe line emission shows a more uniform distribution 
along the plane, with no apparent enhancements to the northeast of Sgr A*. 
This feature suggests that the young star-forming and/or SNe activities have 
less effect on the He-like Fe distribution than they do on the neutral Fe
line emission. The hot Fe line emission might rather be caused by the hot 
ISM trapped and then heated by the strong Galactic center magnetic fields.

The spectral characteristics of the strong neutral Fe line emitting features
to the northeast of Sgr A* are plausibly described by the XRN model. 
We identify dense molecular counterparts for these Fe knots, which is 
strong support for the hypothesized XRN origin.  If past activity of 
Sgr A* is the primary external illuminating source, the required 
luminosities are $L_X$ $\sim$ 10$^{35}$$-$10$^{36}$ ergs s$^{-1}$ with 
durations of at least a few years. 
Alternatively, nonthermally-accelerated electrons by SN shocks penetrating 
back into dense, cold supernova ejecta may also explain the observed neutral 
Fe line features. The presence of active star-forming regions and the
dense molecular clouds near the neutral Fe features are supportive of this
scenario. Unambiguous discriminations between these models are however 
difficult with the current data.  Follow-up observations with the current 
and upcoming observatories may be helpful for a detailed understanding of 
the origins for these features. 

\acknowledgments
{The authors thank M. Tsuboi for providing the CS data. S.P. thank A. Bykov 
for the discussion on the supernova ejecta model for the neutral Fe line
emission. This work has been supported in parts by NASA contract 
NAS8-39073, NAS8-01128 for the {\it Chandra X-Ray Observatory}.}

\clearpage

\begin{deluxetable}{ccc}
\footnotesize
\tablecaption{Chandra/ACIS Observation log
\label{tbl:tab1}}
\tablewidth{0pt}
\tablehead{\colhead{ObsID} & \colhead{Date} & \colhead{Exposure (ks)}}
\startdata
0242 & 1999-09-21 & 41 \\
1561a & 2000-10-26 & 36 \\
1561b & 2001-07-14 & 14 \\
2943 & 2002-05-22 & 35 \\
2951 & 2002-02-19 & 12 \\
2952 & 2002-03-23 & 12 \\
2953 & 2002-04-19 & 12 \\
2954 & 2002-05-07 & 12 \\
3392 & 2002-05-25 & 167 \\
3393 & 2002-05-28 & 158 \\
3663 & 2002-05-24 & 38 \\
3665 & 2002-06-03 & 90 \\
\enddata
\end{deluxetable}

\begin{deluxetable}{cccc}
\footnotesize
\tablecaption{Energy Bands used for Generating the Equivalent Width
Images.
\label{tbl:tab2}}
\tablewidth{0pt}
\tablehead{ \colhead{Elements} & \colhead{Line} &
\colhead{Low\tablenotemark{a}} & \colhead{High\tablenotemark{a}} \\
 & \colhead{(eV)} & \colhead{(eV)} & \colhead{(eV)} }
\startdata
Si (He$\alpha$) & 1700 $-$ 1900 & 1550 $-$ 1640 & 2100 $-$ 2150 \\
S (He$\alpha$ + Ly$\alpha$) & 2350 $-$ 2700 & 2100 $-$ 2150 & 2730 $-$ 2810 \\
Fe (neutral) & 6250 $-$ 6500 & 5000 $-$ 6100 & 7150 $-$ 7300 \\
Fe (He$\alpha$) & 6550 $-$ 6800 & 5000 $-$ 6100 & 7150 $-$ 7300 \\
\enddata
\tablenotetext{a}{The high and low energy bands around the
selected line energies used to estimate the underlying continuum.}

\end{deluxetable}

\begin{deluxetable}{cccccccc}
\footnotesize
\tablecaption{Results of Spectral Fittings
\label{tbl:tab3}}
\tablewidth{0pt}
\tablehead{\colhead{Name} & \colhead{$\Gamma$/$kT$} & \colhead{$N_H$} &
\colhead{Fe Line} & \colhead{EW} & \colhead{Flux\tablenotemark{a}} & 
\colhead{$L_X$\tablenotemark{a}} & \colhead{$\chi^{2}$/$\nu$} \\
     & \colhead{- /(keV)} & \colhead{(10$^{22}$ cm$^{-2}$)} & 
\colhead{Center (keV)} & \colhead{(keV)} & & &} 
\startdata
1 & 1.99$^{+1.05}_{-0.51}$/ - & 32.9$^{+4.0}_{-4.2}$ & 6.40$^{+0.02}_{-0.02}$ &
1.19$^{+0.10}_{-0.10}$ & 5.30$\pm$0.06 & 1.3 & 52.9/51 \\
2 & 3.12$^{+0.80}_{-0.90}$/ - & 36.8$^{+9.6}_{-13.8}$ & 6.40$^{+0.02}_{-0.02}$ &
1.03$^{+0.37}_{-0.23}$ & 5.01$\pm$0.05 & 2.4 & 33.7/44 \\
3 & 1.77$^{+0.09}_{-0.82}$/0.31$^{+0.59}_{-0.08}$ & 15.8$^{+4.1}_{-2.6}$ &
6.39$^{+0.01}_{-0.01}$ & 1.29$^{+0.10}_{-0.10}$ & 5.84$\pm$0.06 & 1.5 & 134.7/121 \\
\enddata
\tablenotetext{a}{The 2 $-$ 10 keV band X-ray flux in the units of 
10$^{-13}$ ergs s$^{-1}$ cm$^{-2}$.} \\
\tablenotetext{b}{The 2 $-$ 10 keV band unabsorbed X-ray luminosity in the 
units of 10$^{34}$ ergs s$^{-1}$.}
\end{deluxetable}

\end{document}